# Determination of five-parameter grain boundary characteristics in nanocrystalline Ni-W by Scanning Precession Electron Diffraction Tomography

Patrick Harrison[1*], Saurabh Mohan Das[2], William Goncalves[3], Alessandra da Silva[4], Xinren Chen[2], Nicola Viganò[5,6], Christian H. Liebscher[2], Wolfgang Ludwig[3,5], Xuyang Zhou[2], Edgar F. Rauch[1*]

[1] Univ. Grenoble Alpes, CNRS, Grenoble INP, SIMAP, F-38000 Grenoble, France

[2] Max-Planck-Institut for Sustainable Materials (Max-Planck-Institut für Eisenforschung), Max-Planck-Strasse 1, 40237 Düsseldorf, Germany

[3] Univ Lyon,CNRS, INSA Lyon, Université Claude Bernard Lyon 1, MATEIS, UMR5510, 69621 Villeurbanne, France

[4] Paul-Drude-Institut für Festkörperelektronik, Leibniz-Institut im Forschungsverbund Berlin e.V., Hausvogteiplatz 5-7, D-10117 Berlin, Germany

[5] ESRF – The European Synchrotron, CS40220, 38043 Grenoble, France

[6] IRIG-MEM, CEA, Université Grenoble Alpes, Grenoble, 38000 France

## Abstract

Determining the full five-parameter grain boundary characteristics from experiments is essential for understanding grain boundaries impact on material properties, improving related models, and designing advanced alloys. However, achieving this is generally challenging, in particular at nanoscale, due to their 3D nature. In our study, we successfully determined the grain boundary characteristics of an annealed nickel-tungsten alloy (NiW) nanocrystalline needle-shaped specimen (tip) containing twins using Scanning Precession Electron Diffraction (SPED) Tomography. The presence of annealing twins in this face-centered cubic (fcc) material gives rise to common reflections in the SPED diffraction patterns, which challenges the reconstruction of orientation-specific virtual dark field (VDF) images required for tomographic reconstruction of the 3D grain shapes. To address this, an automated post-processing step identifies and deselects these shared reflections prior to the reconstruction of the VDF images. Combined with appropriate intensity normalization and projection alignment procedures, this approach enables high-fidelity 3D reconstruction of the individual grains contained in the needle-shaped sample volume. To probe the accuracy of the resulting boundary characteristics, the twin boundary surface normal directions were extracted from the 3D voxelated grain boundary map using a 3D Hough transform. For the sub-set of coherent Σ3 boundaries, the expected {111} grain boundary plane normals were obtained with an angular error of less than 3° for boundary sizes down to 400 nm². This work advances our ability to precisely characterize and understand the complex grain boundaries that govern material properties.

* Corresponding author, email: harrison.p.j@icloud.com, edgar.rauch@grenoble-inp.fr

## Introduction

The properties of polycrystalline materials, including electrical resistivity, mechanical behavior, and corrosion resistance, are affected by the material microstructure [1], which encompasses the materials' crystallography, grain size and shape, and grain boundary characteristics. A grain boundary is characterized in terms of the lattice misorientation across the boundary (3 degrees of freedom - DOF) and the grain boundary plane orientation (2 DOF), and this parameterization is known as the five-parameter grain boundary characteristic distribution (GBCD) [2]. The grain orientation, and hence grain boundary misorientation, may be determined by different techniques, including Electron Backscatter Diffraction (EBSD) [3] or Automated Crystal Orientation Mapping (ACOM) [4] in the Transmission Electron Microscope (TEM). Stereological methods have been used to calculate the GBCD from 2D orientation data [5]. However, these estimate the distribution of grain boundary planes within a sample rather than an individual grain boundary. Determination of an individual grain boundary plane must be calculated from 3-dimensional data and techniques such as Diffraction Contrast Tomography (DCT) [6], 3D-EBSD [7,8] which utilizes Focused Ion Beam (FIB) sectioning, and 3D-OMiTEM [9] which is based on dark-field conical scanning in TEM, have been used towards this end. ACOM has recently been extended to 3-dimensions by combining Scanning Precession Electron Diffraction (SPED) in TEM with a tilt series which allows for both, grain tracking through the tilted reference frames and 3D tomographic reconstruction of individual grains within a polycrystalline sample [10,11] from Virtual Dark Field (VDF) [12] projections of individual grains.

Here, we further develop this technique to determine the grain boundary planes within a Ni-W alloy sample. Multiple Σ3 twin boundaries were observed within the sample. After tomographic reconstruction and assembly into the common 3D sample volume one can extract a 2D grain boundary surface mesh and determine the local grain boundary surface normal directions with respect to the orientation of the adjacent crystal reference. Since Σ3 annealing twin boundaries are known to have {111} surface normals, one can calculate the deviation of the reconstructed grain boundary surface normal from this expected value in order to estimate the accuracy of the 3D grain shape and grain boundary reconstruction process.

## Experimental

An electrodeposited nanocrystalline Ni-14 at% W alloy on a Cu substrate was received from Xtalic Corporation, USA. The electrodeposition route produces a randomly oriented nano-grained material [13]. The sample was annealed in a high vacuum tubular furnace at 600 C for 24 h for remove the amorphous regime in the as-deposited material leading to grain growth with an average grain size of ~70 nm. Precipitation of $Ni_4W$ phases has been observed in this material for annealing times longer than 6h.

SPED experiments were performed using a JEOL JEM-2200FS electron microscope operated at 200 kV. A 0.5° beam precession was supplied by a Digistar hardware unit (NanoMEGAS SPRL). The probe size was measured to be ~2.5 nm. A Fischione Instruments Model 2050 On-Axis Rotation Tomography Holder was used to tilt the sample and projections were acquired with a target tilt step of 15°. Each SPED dataset was acquired at 24 frames per second with a step size of 1.4 nm over a 100 × 276 pixels scan area. The diffracting signal was dominated by the spatial position of the probe despite the observation that the probe size was larger than the scanning step size, this is due to the highly peaked shape of the electron probe distribution.

The general reconstruction methodology for the 3D Scanning Precession Electron Diffraction data has been described in detail elsewhere [10] and was followed here. Briefly, the grain orientations were determined from the SPED data using the Automated Crystal Orientation Mapping (ACOM) tool ASTAR from NanoMEGAS SPRL for each of the tilt series datasets. Each SPED dataset was subjected to 3-pass multi-indexing [14] due to the contribution of overlapping grains in individual diffraction patterns. The dimensionalities of the individual orientation maps were reduced by connecting adjacent pixels with similar crystal orientation (maximum disorientation < 5°) and calculating average statistics, such as orientation and center of mass, these aggregates have the physical interpretation of grains in the sample. Orientation clustering was performed using the Orix library [15]. The precise rotation that couples each successive tilt dataset was then calculated from the set of orientation components. This information allowed grains to be tracked throughout the tilt series and grain orientations in different tilt reference frames to be predicted.

Projections of the individual grains were calculated using a variant of Frozen Template Virtual Dark Field (FTVDF) reconstruction [16], which systematically excludes shared reflections common with other twin related orientation components in the sample volume. These shared reflections can give rise to "ghost images" - i.e. systematic superposition and contamination of the FTVDF projection of a given grain with signal from twin related orientation components which in turn will violate tomographic consistency of the projection data.

Coarse alignment of the tilt series was performed using a cross-correlation technique on the Virtual Bright Field (VBF) projections and subsequent fine alignment was achieved using the IMOD software [17] by using the FTVDF images of individual grains as markers for fiducial alignment. Moreover, intensity normalizations prior to 3D grain shape reconstruction were used to correct for the decay of the electron probe in thicker regions of the tip (division by diffraction frame total intensity, see Supplementary Information Fig. S1) and to force the summed intensity in the individual FTVDF grain projection images to a common value for all tilts. Each grain was reconstructed individually into a common sample volume through 15 iterations of the Simultaneous Iterative Reconstruction Technique (SIRT) using the ASTRA toolbox [18]. The intensity values of these individual grain reconstructions were re-normalized such that voxels belonging to central regions of the different grains exhibit similar gray values. Finally, a labelled sample volume (integer numbers corresponding to grain IDs) was created using a max-pooling algorithm where each of the voxels was assigned to the grain ID with the highest reconstructed intensity at the corresponding location.

## Results

*Recognizing twinned grains*

The orientation map calculated from the 80° tilt dataset is shown in Figure 1.a
. Further analysis of misorientation between neighboring grains revealed multiple Σ3 twin boundary relationships, defined as a 60° rotation around the crystal <111> axis, as can be seen in Figure 1.b-e, including a large multiply twinned grain highlighted by line profile 1. The advantage of dealing with twins is that their crystallographic characteristics are perfectly known and can be used to quantify the accuracy of the 3D reconstruction with no a priori knowledge. A challenge is that twinned grains have very similar diffracting signature for certain orientations. In particular, if the (111) common plane is parallel to the electron beam, there is no diffracting spot unique to only one of the grains in the Zero-Order Laue Zone (ZOLZ). Statistically this exact configuration is very improbable, but for orientations close to this configuration there are still numerous common diffraction spots which in turn have a detrimental effect on the reconstruction procedure

Virtual Dark Field (VDF) reconstruction techniques are used to generate projections of individual grains from the diffraction data. To this end, the FTVDF technique, which uses the simulated Bragg peak positions derived from calculated diffraction templates, has been shown to produce improved representations of the grain contours and can be applied in an automated way, reducing processing time and operator error [10,16]. With such a procedure, a clear separation of the twinned grains using FTVDF will be hard for the orientations mentioned above due to the existence of numerous common reciprocal lattice points between the two crystal orientations. As a result, there is minimal contrast in the FTVDF reconstructed images between the twinned grains, as shown in Figure 2, which is detrimental to tomographic reconstruction.

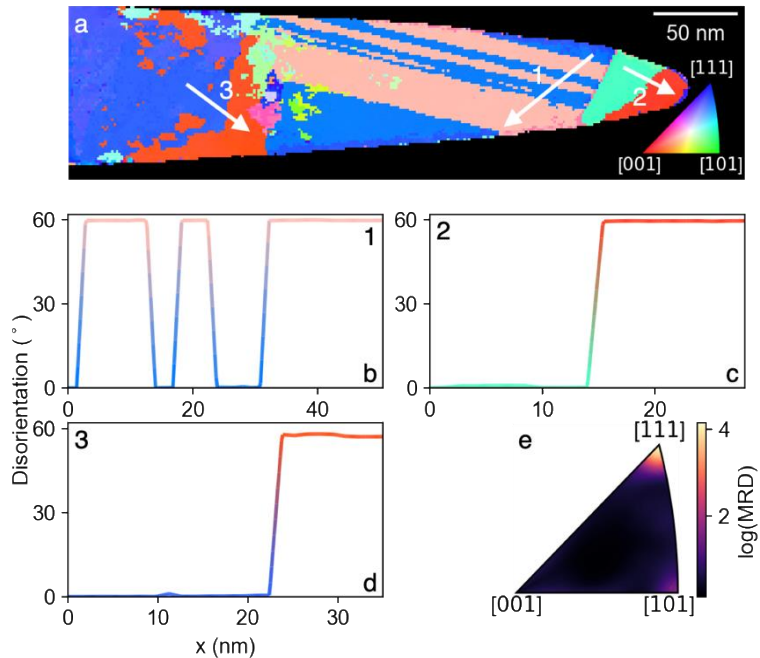

*Figure 1: Occurrence of Σ3 twinned grains in the NiW sample. (a) Orientation map in IPF coloring projected along tilt axis (color code inset) calculated from the 80° projection. The white arrows (1-3) correspond to disorientation line profiles (b-d). The disorientation in each case was computed with respect to the first orientation data point on each line. (e) The inverse pole density distribution of the misorientation axis of all grain boundaries within the sample accumulated from the 2D orientation maps of all tilt datasets. Units are in multiples of random distribution (MRD).*

To overcome this contrast issue, whilst maintaining the required level of automation for reconstructing multiple grains within the sample at multiple tilts, we have developed algorithms to account for the occurrence of common reciprocal lattice points when the diffraction templates are generated. During the diffraction template simulation, the set of valid reciprocal lattice points for a given structure is computed. At this point, a second set of reciprocal lattice points is computed by rotating the initial set around the crystal misorientation axis by the misorientation angle, and the overlapping reciprocal lattice points are removed from the initial set. An example of this procedure is shown in Figure 2.a for Σ3 twinned grains, where the red and cyan points represent unique points in the two reciprocal lattices and the black points are common between them. When using the black reciprocal lattice points for VDF reconstruction, there will be significantly less observable contrast between the grains. An example demonstrating the contrast improvement in the FTVDF reconstructions of twinned grains when the common points are removed is shown in Figure 2.b. A drawback of this technique is that each FTVDF is reconstructed using less total diffraction signal, however the improved contrast means that previously convoluted grains may now be independently resolved.

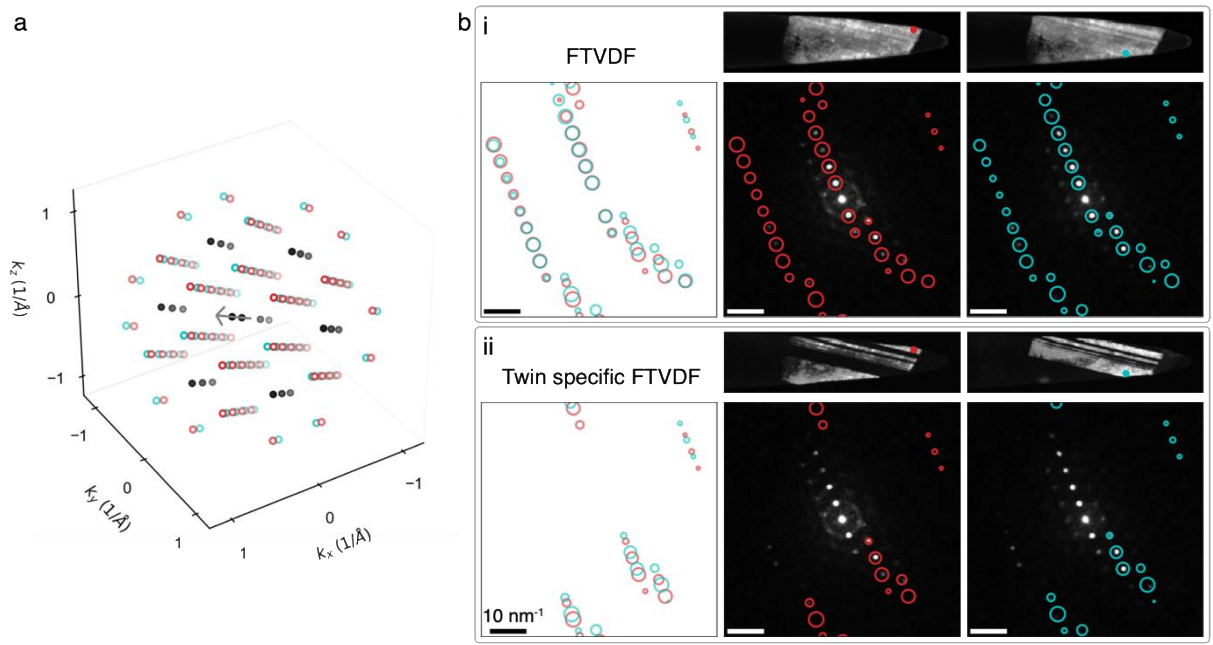

*Figure 2: FTVDF reconstruction using misorientation specific virtual apertures. (a) Simulated reciprocal lattice points for an Ni crystal. Two reciprocal lattices are simulated, red and cyan, which differ by 60° rotation around the crystal [111] axis (shown as a grey arrow), this is the orientation relationship for Σ3 twins. These two reciprocal lattices share common points in reciprocal space, which are shown in black. (b) Diffraction pattern recorded from two Σ3 twinned grains (positions denoted by points in FTVDF reconstructed images) are shown. The orientation measured by ACOM is used to simulate the virtual apertures in the normal FTVDF calculation (i) (aperture locations for the two twinned orientations are shown as red and cyan, respectively), however there is significant overlap between aperture positions. As a result, there is limited contrast between the twinned grains in the FTVDF reconstructed images (top). (ii) The reciprocal lattice for each grain orientation was simulated and overlapping reciprocal lattice points were removed. The resulting reduced set of reciprocal lattice points were used to generate kinematical diffraction templates. By using these twin specific FTVDF apertures, the contrast between the twinned grains is significantly improved.*

### 3D reconstruction of the Ni-W alloy tip

The procedure was applied to the Ni-14 at% W alloy tip. Due to some technical issues, the 15° tilt increment was never exact and for one particular scan pretty far for the expected tilt angle. For SPED scans this is not a problem because the reconstruction uses the tilt angles and axis that are determined from the set of computed orientations. The only concern is that 11 rather than 12 tilts were properly meaningful for the reconstruction.

The sample contains 5 twinned Ni grains (fcc, lattice parameter 3.52 Å) and a large particle of $Ni_4W$ phase (Tetragonal, I4/m) leading to 11 components, see Figure 3.a. In the present context, a component is a volume that is characterized by one set of crystallographic parameters. Therefore, it is either a crystal (e.g. the $Ni_4W$ particle) or one of the two entities contained in a twinned grain. Each component was reconstructed separately and shares specific interfaces with its neighbors. Three of the grains contain several twins that share the same orientation. In other words, these grains are composed of two components but enclose more than one twin plane (e.g.: the twin system marked 1 in Figure 3.b).

### Twin boundaries
The capability of the procedure to properly recognize interfaces was quantified by comparing the known orientation of twin boundaries to the orientation of the same boundaries deduced solely from the reconstruction.

Twin boundaries derived from crystallographic orientations: coherent twin boundaries in fcc materials are planes normal to a <111> vector. These vectors are easily extracted from the crystal orientation data by computing the common axis between the parent crystal and the twin. During tilting, these vectors move around the tilt axis by describing cones in the laboratory reference frame. In the pole figure of Figure 3.c, these cones are represented by colored circles, each being associated to a specific twinned grain. In the same figure, the circular markers denote the location of the calculated twin plane normal for the successive tilt angles.

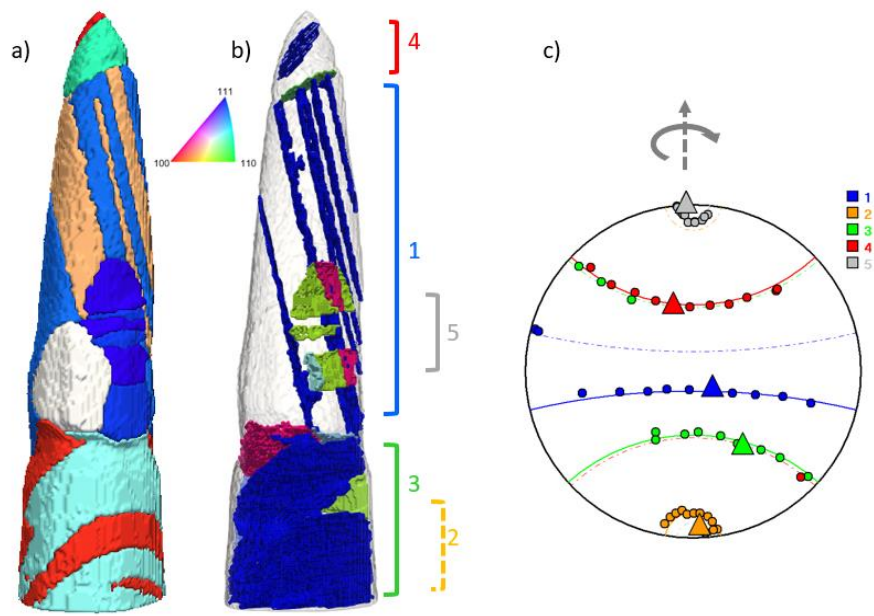

Figure 3: The 3D reconstructed Ni-W tip and twin plane location. (a) the Ni-W nanocrystalline sample in which each grain is colored according to its average orientation projected along the vertical axis (color code inserted); the large Ni4W particle is highlighted in white. (b) grain boundary habit planes for the Ni nano grains solely colored with respect to the boundary normal (same color code as (a)). The dominant blue color is related to the numerous {111} twin planes appearing in this material. One of them (number 2) is hidden by the front grains (c) pole figure for the Σ3 twinned grains normal axis. The colors represent the different twinned grains located by the brackets in (b). Colored markers represent the {111} axis poles as calculated from the orientation data for every successive tilt around the vertical axis. The corresponding large triangles are the estimated twin plane orientations as measured from the reconstructed volume at 0° tilt angle.

Twin boundaries derived from the 3D reconstruction: the reconstructed volumes are initially delimited by isosurfaces defined as the collection of voxels belonging to this crystal and sharing the same intensity as computed through the (Simultaneous Iterative Reconstruction Technique) SIRT algorithm [19]. A simple juxtaposition of these volumes would lead to holes or overlapping areas depending on the thresholds used to define the isosurfaces. To locate the boundaries, these surfaces can no longer be identified by constant values, instead a max-pooling type approach is used. It attributes to each voxel the component with the highest intensity. The resulting volumes are by construction limited by the adjacent components. Consequently, the boundary between components i and j are defined as the collection of voxels belonging to one of the components and touching the other one. It is important to note that in this process there is no consideration of any crystallographic orientation.

The boundaries of the reconstructed volumes are expected to be planar so a simplified 3D Hough transform has been used to retrieve their orientations (Supplementary Information Fig. S2): the interface voxels are projected as pixels in every slice of the stack of data. These 2D projections of the boundaries contain lines, or segments of lines that are identified with a standard Hough transform. This gives two of the three independent coordinates of the interface normal. The third coordinate corresponds to a displacement of the lines from one slice to the next and is obtained by scanning its value in the full range of possible slope, counteracting the displacement accordingly, and summing the Hough accumulators from all slices in each case (Fig S2). The slope that gives the highest accumulation defines the final coordinate. Note that the accumulation properly works only if the interface is planar. If not, the values would be spread in Hough space and there would be no result. Consequently, the procedure helps to estimate the orientation of the interfaces but also, to some extent, detects their flatness.

*Plane normal accuracy:*

The plane normals derived from the reconstructed volume are shown as large triangles in Figure 3.c for the 0° tilt. They all agree with the expected orientations. Their angular distance from the normal paths – the cones mentioned above - is in average 2.5°. This value is a lower bound for the accuracy. When compared to the specific normals derived from the orientation maps at 0° tilt, this average rises to 4.6°. The latter is comprised of both the error related to the reconstruction and the inaccuracy in the orientation mapping procedure for this particular tilt. It may be considered as an upper bound for the error.

The values are plotted in Figures 4.a and 4.b. It is expected that the accuracy improves with increasing twin plane size, i.e. with more data. And indeed, the largest twin system (n° 1 in Fig. 3) has the best estimate (err: 2.2°). To verify this point, the sizes are computed by considering the number of points that define the interface as the twin plane surfaces scale roughly with this number times the area per pixel. The latter is the square of the scan step size. For the present data the twin plane surfaces range between 400 nm² and 40000 nm². Note that these values are the sum of the surfaces when multiple twin planes appear in a given grain, typically 5 planes for the largest twin system. Figure 4.a demonstrates that there is no such correlation between the size and the angular error. One of the smallest twin planes (~15x15 nm²) exhibits one of the highest accuracies (n° 4 in Fig. 3, error: 2.3°).

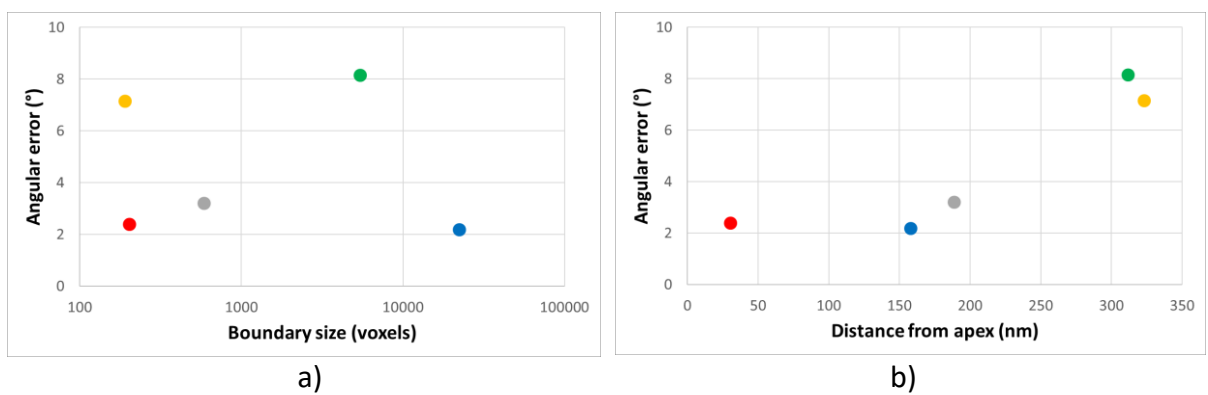

a) b)

*Figure 4: Angular error computed for the five twin systems (a) with respect to the number of points used to measure the plane normal orientation, (b) as a function of their position in the tip. The colors refer to the twin planes identified in figure 3.c*

The related grain is at the apex of the tip which suggests that the position of the boundary in the reconstructed volume has an impact. In Figure 4.b, it appears that there is a correlation between the angular error and the distance from the apex. There are two possible explanations for this trend: firstly, the diameter of the tip increases from approximatively 40 to 80 nm leading to a decrease of the quality of the diffraction patterns acquired in the thicker part of the tip and secondly, the alignment of the tilt datasets is more reliable towards the tip apex as a result there may be errors related to the exact position of the plane at the other side.

In any case, the present exercise demonstrates that the full five parameter description of nanoscale boundaries is achievable with good accuracy through the tilt series of SPED scans described in this work.

*Spatial Resolution*

Further analysis into the spatial resolution of the reconstruction technique outlined in this work is shown in Figure 4, which compares the final indexed reconstruction to the twin-specific FTVDF reconstructed images of the large coherent multiply twinned grain described in previous sections. The region highlighted by the white bracket in Figure 4.b-d reveals a section of the 3D reconstruction indexed to the blue grain, however this region actually contains multiple thin sections that should be indexed to the orange twin as highlighted by white arrows in Figure 4.d. This observation is also available in orientation map shown in Figure 1.a, however it is only partially visible for only one of the twinned sections and the missing information is also not recovered through multi-indexing.

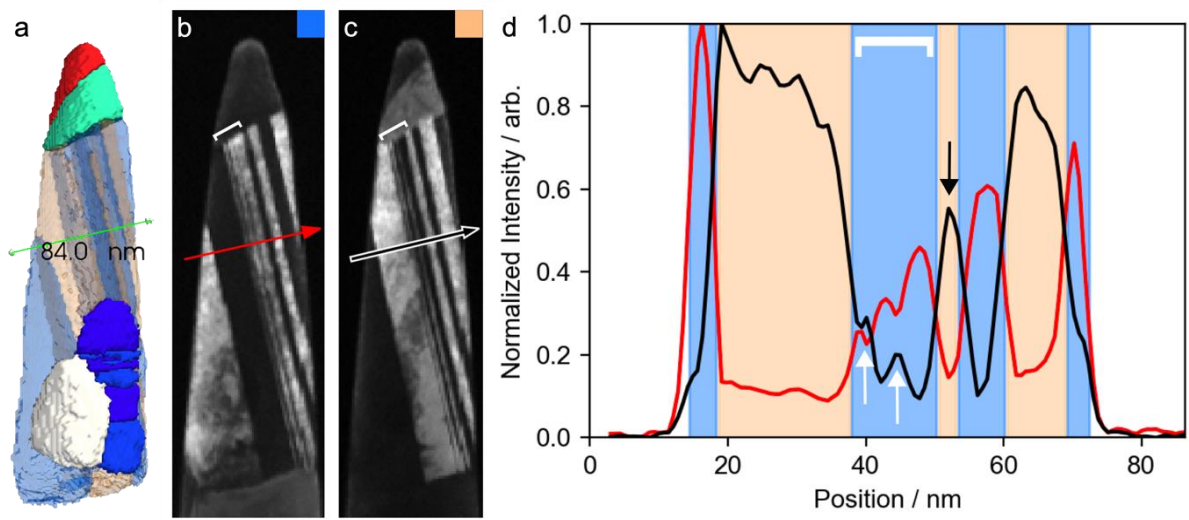

*Figure 4: Limitations reconstructing thin coherent twins. (a) Line profile (green arrow) through multiply twinned grains (blue and orange) in the indexed 3D reconstructed tip. The line profile direction was oriented along the measured twin-plane normal. The colour code of the delimited grain regions is maintained in (b-d). (b, c) Twin-specific FTVDF reconstructed images of the blue and orange grains in (a), respectively, from the 80° tilt dataset. The 2D line profiles corresponding to the 3D line profile in (a) are coloured as red and black in (b, c), respectively, and maintained in (d). (d) Normalized intensity profiles along the lines in (a-c).*

The width of these twins is only 2-3 pixels in the raw FTVDF data, corresponding to ~3.5nm. By contrast the neighbouring twinned section, indicated by the black arrow in Figure 4.d, is correctly reconstructed. With a width of ~7 nm and as the smallest feature reconstructed in this dataset, this defines the upper bound of the spatial resolution. The twins that were not reconstructed have significantly less intensity in the FTVDF projection relative to the neighbouring parent grain when compared to the twinned section that was reconstructed, and this factor contributes significantly to the success of the reconstruction. As discussed in previous sections, whilst the twin-specific FTVDF reconstruction algorithm improves orientation specificity, it does so at the expense of overall signal during the reconstruction process. As a result, further work into improving the reconstruction quality of such small features using the technique outlined in this work is envisaged to include both algorithmic improvements and optimizing experimental conditions, such as minor adjustments to tilt angle or dwell time, to maximize signal for the FTVDF reconstruction.

## Discussion

The present work shows a procedure to determine the 5 macroscopic degrees of freedom of grain boundaries obtained through the present 3D-SPED technique. We use a nanotwinned region as a model system to demonstrate the robustness of the procedure and to elucidate the angular and spatial resolution of the 3D reconstruction. By using the simplified configuration of twin boundaries, it is demonstrated in the present work that such reconstructions are feasible for nanoscaled boundaries. The work also points out that the measurement accuracy is particularly sensitive to the tilt alignments and distortion correction so that dedicated procedures are under investigation for making further improvement to the quality of grain boundary characterizations.

At the current stage of development, the full process is time consuming and involves many semi-automated or manual steps in the reconstruction. In particular, the pattern collection procedure is long, which is limited by the maximum frame rate obtainable with the employed scintillator coupled CMOS detector. The implementation of faster and more sensitive cameras equipped with a direct electron detector will transform this workflow in a quicker procedure from currently ~20 mins per tilt step to less than one minute when using the same scan grid as employed here. This would not only reduce the impact of sample drift and related artifacts during the acquisition, but also lower the total electron dose deposited onto the sample, which becomes relevant when reconstructing beam sensitive samples. It would also allow to use finer tilt increments and more projections for the tomographic reconstruction, which would lead to an increase in total dataset size and may limit post acquisition data analysis. Dedicated data analysis workflows (graphical interfaces, scripts and Jupyter notebooks) are under development, guiding the operator through the different steps of the analysis procedure and to promote future routine use of the technique for the analysis of needle-shaped specimen and in particular atom probe tips for 3D correlative experiments in combination with APT.

## Conclusions

The 3D grain microstructure and associated grain boundary characteristics of an annealed nickel-tungsten alloy (NiW) polycrystalline tip have been determined from a tomographic reconstruction using tilt series of scanning precession electron diffraction data. The grain shapes and twin boundary surface normals have been reconstructed with no a priori information concerning their planar shape and orientation. It is shown that the reconstructed twin boundary normal directions are within a few degrees from the expected <111> directions of the adjacent crystal lattices. This demonstrates that boundary characteristics may be retrieved with reasonable accuracy using tomographic reconstruction through virtual dark field projections and opens the door for future works investigating nanoscale grain boundaries in transmission electron microscopy.

## Acknowledgements

P.H., W.L, W.G. and E.F.R. would like to recognize funding from the Agence Nationale de la Recherche (grant no. ANR-19-CE42–0017). X.Z. and S. M.D acknowledge funding by the German Research Foundation (DFG) via project HE 7225/11–1. X. Z. is supported by the Alexander von Humboldt Stiftung.

## CRediT authorship contribution statement

**Patrick Harrison**: Conceptualization, Investigation, Software, Writing – original draft, **Saurabh Mohan Das**: Investigation, Methodology, **William Goncalves:** Software, methodology, **Alessandra da Silva:** Methodology, **Xinren Chen:** Software, Methodology, **Nicola Viganò:** Methodology, **Christian H. Liebscher:** Writing – review & editing **Wolfgang Ludwig:** Conceptualization, Supervision, Methodology, Writing – review & editing, Funding acquisition, **Xuyang Zhou:** Supervision, Writing – review & editing, Funding acquisition, **Edgar F. Rauch:** Conceptualization, Supervision, Methodology, Writing – original draft, Funding acquisition.

# Supplementary Information

**Determination of five-parameter grain boundary characteristics in nanocrystalline Ni-W by Scanning Precession Electron Diffraction Tomography**

Patrick Harrison1*, Saurabh Mohan Das2, William Goncalves3, Alessandra da Silva4, Xinren Chen2, Nicola Viganò5,6, Christian H. Liebscher2, Wolfgang Ludwig3,5, Xuyang Zhou2, Edgar F. Rauch1*

1 Univ. Grenoble Alpes, CNRS, Grenoble INP, SIMAP, F-38000 Grenoble, France
2 Max-Planck-Institut for Sustainable Materials (Max-Planck-Institut für Eisenforschung), Max-Planck-Strasse 1, 40237 Düsseldorf, Germany
3 Univ Lyon,CNRS, INSA Lyon, Université Claude Bernard Lyon 1, MATEIS, UMR5510, 69621 Villeurbanne, France
4 Paul-Drude-Institut für Festkörperelektronik, Leibniz-Institut im Forschungsverbund Berlin e.V., Hausvogteiplatz 5-7, D-10117 Berlin, Germany
5 ESRF – The European Synchrotron, CS40220, 38043 Grenoble, France
6 IRIG-MEM, CEA, Université Grenoble Alpes, Grenoble, 38000, France

*Frame Total Image*

The Frame total image sums the transmitted and diffracted beam intensities and is used to normalize the FTVDF images of every component in order to compensate for the local thickness of the sample.

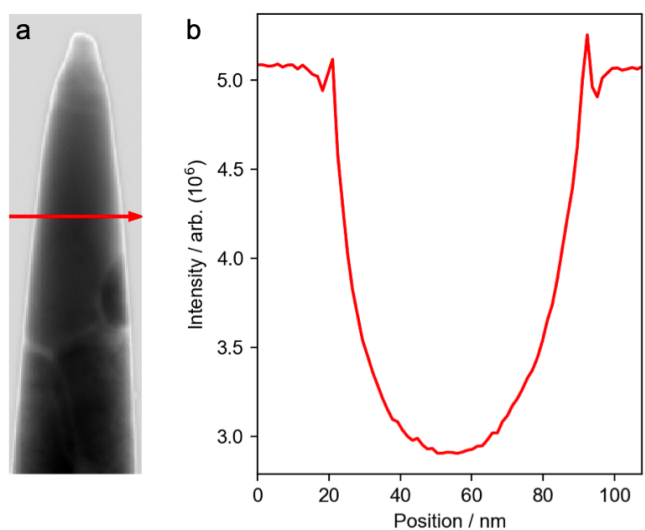

*Figure S1: (a) The diffraction frame total image for the 0° dataset, reconstructed by summing the diffraction frame intensities for each probe position. (b) The intensity line profile taken across the red arrow in (a). The profile is in good agreement with the expected projection of an object with a circular cross-section.*

*3D Hough transform*

The twin plane orientation is deduced from a 3D Hough transform that cumulates the 2D Hough signal of the reconstructed plane over all slices. The 'Line inclination' measures the slope of the lines appearing in Figure S2.b (-6° in the example shown) and provides two of the directional cosines of the plane normal. The third directional cosine is derived from the vertical shift required between successive slices to maximize the signal in the Hough space (no shift in Fig S2.a, Δy/Δz =0.21 i.e.: 12° in Fig.S2.c). The reference frame is adapted to the plane orientation: typically for planes nearly perpendicular to the tip length, the same construction is performed in the (z,y,-x) reference frame instead of the (x,y,z) one.

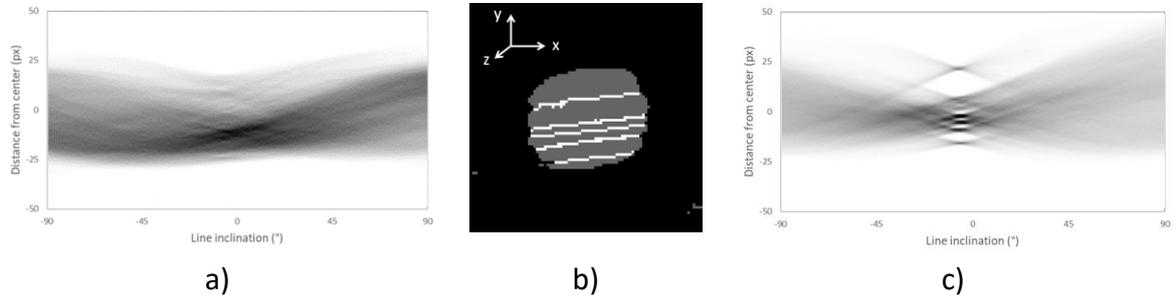

a) b) c)

*Figure S2: The 3D Hough transform construction. The segment of lines appearing in each slice (b) are projected in the usual angle-distance plot (a and c). When cumulated over all the slices the signal is spread in the Hough space (a). By adapting the relative position of each slice with a constant displacement along y per slice (for the example show) the accumulation is maximized.*